\documentstyle[11pt]{article}
\def\be{\begin{equation}}\def\ee{\end{equation}}\def\ni{\noindent}
\def\bb{\begin{thebibliography}{99}\small\addtolength{\itemsep}{-6pt}}
\def\eb{\end{thebibliography}}\newcommand{\tr}{{\rm tr}\,}
\renewcommand{\det}{{\rm det}}\newcommand{\eqr}[1]{(\ref{#1})}
\def\cD{{\cal D}}\def\cA{{\cal A}}\def\cS{{\cal S}}
\def\e{~\hbox{e}}\def\l{\lambda}\def\lo{\l_o}\def\da{\dagger}
\def\d{\partial}\def\D{\Delta}\def\la{\langle}\def\ra{\rangle}
\def\f{{\rm f}}
\newcommand{\N}{{\scriptscriptstyle N}}
\newcommand{\R}{{\scriptscriptstyle R}}
\begin{document}
\begin{flushright}hep-th/9703142\\OUTP-97-13P\\March 1997\end{flushright}
\vskip.5cm
\begin{center}{\Large\bf Exactly Soluble QCD and Confinement of Quarks}\\
\vskip.5cm {\large\bf B. Rusakov}\\\vskip.1cm{\em Theoretical Physics, 
Oxford University\\1 Keble Road, Oxford, OX1 3NP, U.K.}\vskip.5cm
\end{center}{\small 
An exactly soluble non-perturbative model of the pure gauge QCD is derived
as a weak coupling limit of the lattice theory in plaquette formulation
\cite{1}. The model represents QCD as a theory of the weakly interacting 
field strength fluxes. The area law behavior of the Wilson loop average is
a direct result of this representation: the total flux through macroscopic
loop is the additive (due to the weakness of the interaction) function of
the elementary fluxes. The compactness of the gauge group is shown to be
the factor which prevents the elementary fluxes contributions from
cancellation. There is no area law in the non-compact theory.}

\section{Introduction.}
It is well understood that the confinement of quarks can be described by
the non-perturbative QCD only. The discovery of asymptotic
freedom \cite{GW} within perturbative approach to the Yang-Mills theory 
unambiguously peaks QCD as a genuine model of strong interaction. However,
the growth of the effective coupling constant towards infrared region,
which is the opposite prediction of asymptotic freedom, makes perturbative 
theory unreliable in infrared. Since we cannot make sense of the path
integrals except the gaussian case, the derivation of the non-perturbative
solution of the Yang-Mills model, which is equivalent to the practically 
impossible summation of all Feynman diagrams, cannot be considered as a
realistic goal.

The unique known opportunity to solve the problem non-perturbatively is
provided by the Wilson lattice model \cite{Wils} which is exactly
soluble\footnote{At least, numerically.} for any coupling constant (i.e.,
non-perturbatively) on any finite lattice. In two-dimensions, for example,
the lattice model easily gives an exact solution \cite{Mig,90} in 
continuum. In $D>2$ the problem is much more complicated. However, since
the lattice model already contains all of the non-perturbative
information, the remaining problem of taking continuum limit of the
lattice results, in spite of being also a non-trivial task, is essentially
more attractive.

It is very important, in this respect, to formulate the lattice model in
the most adequate terms. We shall argue that such formulation is the 
plaquette formulation performed for $D=3$ in \cite{1} (this was given in
the form different from that of the pioneering works \cite{HB}). This
formulation is the starting point for present analysis, and will be
briefly reviewed in the next section. The motivation (given also in
\cite{1}) is as follows.

The partition function of the Wilson pure gauge model in $D$ dimensions
can be written as\be Z=\int\prod_ldU_l
\prod_pdU_p\e^{\frac{N}{\lo}\tr(U_p+U_p^\da)}\delta\Big(U_p,\prod_{l
\in p}U_l\Big)\;\;,\label{WZ}\ee where $\lo$ is the bare coupling constant, 
$l$ and $p$ denote links and plaquettes of $D$-dimensional lattice, $U_l$
is the unitary matrix ($U(N)$ or $SU(N)$) attached to $l$-th link. The
products of $U_l$'s are ordered according to the geometrical order of
links.
 
The gauge-invariant $\delta$-function is\be\delta(U,V)=\sum_r\chi_r(U^\da)
\chi_r(V)\;,\label{d}\ee where $r$ is an irreducible representation with
the character $\chi_r(U)$ and dimension $d_r=\chi_r(I)$.

The only non-zero observables in this model are invariant ordered products
of link variables along {\em closed} loops, such as 
\be W(C)=\la\frac{1}{N}\tr\prod_{l\in C}U_l\ra\label{wl}\ee ($C$ is the
closed contour). It is understood for a long time that the plaquette 
variables \cite{HB,MM} are relevant to solution of the quark confinement
problem \cite{Wils}. As argued in \cite{Wils}, the area-law behavior of
\eqr{wl},
\be W(C)\sim\e^{-\sigma\cA}\;,\label{arl}\ee where $\cA$ is the area of
the minimal surface $\cS$ bounded by $C$, and $\sigma$ is the
positive parameter (string tension), means {\em confinement of quarks},
since it corresponds to the linear potential between colored objects.

The practical use of the idea of dealing with only the loop variables has
been first explored by Halpern and Batrouni \cite{HB} in their field
strength and plaquette variable formulation of gauge theory and by
Makeenko and Migdal \cite{MM} in relation to the loop equation (see also
\cite{Mig83} and references therein). However, in spite of all efforts and
the progress achieved in these directions, the solution to the confinement
problem has not been found.

We follow here a variant \cite{1} of approach \cite{HB} which is based on
such formulation of the model in which the only independent variables of
the model are the plaquette matrices $U_p$. This reformulation requires
calculation of the integral
\be\int\prod_ldU_l\prod_p\delta\Big(U_p,\prod_{l\in p}U_l\Big)\label{int}
\ee in \eqr{WZ}, which is responsible for the interaction between
plaquettes. 

Physically, the plaquette matrices are nothing but extended to the compact
group manifold field strength fluxes (just like link matrices are the
group-extended vector-potentials). Confinement, or area law, emerges as
the consequence of the weakness of the interaction between fluxes -- the
total flux through macroscopic loop is the additive function of the
elementary fluxes (plaquettes), i.e. proportional to the minimal area.
The additiveness of elementary fluxes is the necessary but not sufficient 
condition for the confinement: one has to find a reason why these
contributions are accumulated and not canceled in the total flux. We will
find that the reason is the compactness of the gauge group. We will also
show that the additiveness is indeed emerges in the $\lo\to 0$ limit
(which is the only interesting limit as it corresponds to the continuum).
In this limit, as it is easy to see from \eqr{WZ}, the leading
contribution comes from the saddle point $U_p=I$, and indeed a
certain factorization property has to emerge. This leads to a new model
(described below) where the average \eqr{wl} takes the form
\footnote{Such average, called ``filled Wilson loop", first considered
in \cite{K} in a different model.}:
\be W(C)=\la\prod_{p\in\cS}\frac{\tr U_p}{N}\ra\label{wlp}\ee 
where $\cS$ is arbitrary surface bounded by $C$. In the model we will
present, this average is surface-invariant. We emphasize that the average
\eqr{wlp} considered in the model \eqr{WZ} is {\em not} surface-invariant,
except abelian case (where \eqr{wlp} coincides with \eqr{wl}).

The providers of required surface-invariance are the $\delta$-functions of
the integral \eqr{int}. In paper \cite{1} this integral has been computed
exactly (for $D=3$), thus, providing the starting point for present
analysis. 

The paper organized as follows. In Section 2, we briefly repeat the 
plaquette formulation of \cite{1} and write the result as a statistical
model. In Section 3, we suggest factorization of characters in the weak
coupling limit of the lattice theory and derive the effective model of QCD
in arbitrary $D$. In Section 4, we discuss compactness and suggest a new
method to compute the averages in the compact theory. We compute the
Wilson loop average and string tension in general form and give a new
criterion for confinement. We also show that in the mean field
approximation our solution indeed gives confinement and reproduces the
known results in both strong coupling and perturbative regimes.  
In Section 5, we summarize and discuss the results. We claim that our
results mean the rigorous proof of confinement in QCD. In Appendix, we
give some useful group-theoretical formulas relevant to the heat kernel
analysis.

\section{Lattice QCD$_3$ as a statistical model.}
After taking integral \eqr{int} and substituting it into \eqr{WZ} one
obtains \cite{1} the model which is defined on the two-dimensional
infinite genus lattice formed by hexagons $h$ as shown in Fig.\ref{fig1}. 

\def\le{\left(\begin{array}{ll}&\\&\end{array}\right.}
\def\ri{\left.\begin{array}{ll}&\\&\end{array}\right)}
\def\rob{\begin{picture}(15,5)(-15,-5)
\put(14.5,-1.7){\makebox(0,0){.}}\put(-14.5,-1.7){\makebox(0,0){.}}
\put(14,-2.1){\makebox(0,0){.}}\put(-14,-2.1){\makebox(0,0){.}}
\put(13.5,-2.4){\makebox(0,0){.}}\put(-13.5,-2.4){\makebox(0,0){.}}
\put(13,-2.6){\makebox(0,0){.}}\put(-13,-2.6){\makebox(0,0){.}}
\put(12,-2.85){\makebox(0,0){.}}\put(-12,-2.85){\makebox(0,0){.}}
\put(11,-3.05){\makebox(0,0){.}}\put(-11,-3.05){\makebox(0,0){.}}
\put(10,-3.2){\makebox(0,0){.}}\put(-10,-3.2){\makebox(0,0){.}}
\put(9,-3.33){\makebox(0,0){.}}\put(-9,-3.33){\makebox(0,0){.}}
\put(8,-3.45){\makebox(0,0){.}}\put(-8,-3.45){\makebox(0,0){.}}
\put(7,-3.55){\makebox(0,0){.}}\put(-7,-3.55){\makebox(0,0){.}}
\put(6,-3.63){\makebox(0,0){.}}\put(-6,-3.63){\makebox(0,0){.}}
\put(5,-3.70){\makebox(0,0){.}}\put(-5,-3.70){\makebox(0,0){.}}
\put(4,-3.76){\makebox(0,0){.}}\put(-4,-3.76){\makebox(0,0){.}}
\put(3,-3.81){\makebox(0,0){.}}\put(-3,-3.81){\makebox(0,0){.}}
\put(2,-3.85){\makebox(0,0){.}}\put(-2,-3.85){\makebox(0,0){.}}
\put(1,-3.88){\makebox(0,0){.}}\put(-1,-3.88){\makebox(0,0){.}}
\put(0.5,-3.9){\makebox(0,0){.}}\put(-0.5,-3.9){\makebox(0,0){.}}
\put(0.2,-3.91){\makebox(0,0){.}}\put(-0.2,-3.91){\makebox(0,0){.}}
\end{picture}}

\def\rot{\begin{picture}(15,5)(-15,-5)
\put(14.5,1.7){\makebox(0,0){.}}\put(-14.5,1.7){\makebox(0,0){.}}
\put(14,2.1){\makebox(0,0){.}}\put(-14,2.1){\makebox(0,0){.}}
\put(13.5,2.4){\makebox(0,0){.}}\put(-13.5,2.4){\makebox(0,0){.}}
\put(13,2.6){\makebox(0,0){.}}\put(-13,2.6){\makebox(0,0){.}}
\put(12,2.85){\makebox(0,0){.}}\put(-12,2.85){\makebox(0,0){.}}
\put(11,3.05){\makebox(0,0){.}}\put(-11,3.05){\makebox(0,0){.}}
\put(10,3.2){\makebox(0,0){.}}\put(-10,3.2){\makebox(0,0){.}}
\put(9,3.33){\makebox(0,0){.}}\put(-9,3.33){\makebox(0,0){.}}
\put(8,3.45){\makebox(0,0){.}}\put(-8,3.45){\makebox(0,0){.}}
\put(7,3.55){\makebox(0,0){.}}\put(-7,3.55){\makebox(0,0){.}}
\put(6,3.63){\makebox(0,0){.}}\put(-6,3.63){\makebox(0,0){.}}
\put(5,3.70){\makebox(0,0){.}}\put(-5,3.70){\makebox(0,0){.}}
\put(4,3.76){\makebox(0,0){.}}\put(-4,3.76){\makebox(0,0){.}}
\put(3,3.81){\makebox(0,0){.}}\put(-3,3.81){\makebox(0,0){.}}
\put(2,3.85){\makebox(0,0){.}}\put(-2,3.85){\makebox(0,0){.}}
\put(1,3.88){\makebox(0,0){.}}\put(-1,3.88){\makebox(0,0){.}}
\put(0.5,3.9){\makebox(0,0){.}}\put(-0.5,3.9){\makebox(0,0){.}}
\put(0.2,3.91){\makebox(0,0){.}}\put(-0.2,3.91){\makebox(0,0){.}}
\end{picture}}

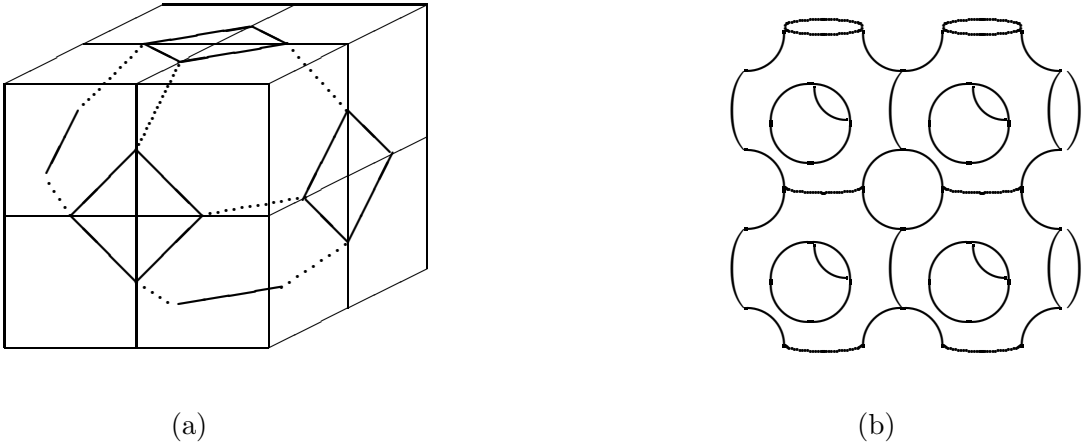
\begin{figure}[h]
\centering\begin{picture}(200,170)(-100,-50)\thinlines
\put(-200,-20){\line(1,0){100}}\put(-200,-20){\line(0,1){100}}
\put(-200, 80){\line(1,0){100}}\put(-100,-20){\line(0,1){100}}
\put(-150,-20){\line(0,1){100}}\put(-200, 30){\line(1,0){100}}
\put(-140,110){\line(1,0){100}}\put(-40,10){\line(0,1){100}}
\put(-170,95){\line(1,0){100}}\put(-70,-5){\line(0,1){100}}
\put(-100,-20){\line(2,1){60}}\put(-200,80){\line(2,1){60}}
\put(-100,80){\line(2,1){60}}\put(-150,80){\line(2,1){60}}
\put(-100,30){\line(2,1){60}}
\thicklines
\put(-150,5){\line(1,1){25}}\put(-150,5){\line(-1,1){25}}
\put(-150,55){\line(1,-1){25}}\put(-150,55){\line(-1,-1){25}}
\multiput(-150,55)(1,2){17}{\makebox(0,0){.}}
\multiput(-150,5)(3,-2){5}{\makebox(0,0){.}}
\put(-148,94){\makebox(0,0){.}}\put(-150,92){\makebox(0,0){.}}
\put(-152,90){\makebox(0,0){.}}\put(-154,88){\makebox(0,0){.}}
\put(-156,86){\makebox(0,0){.}}\put(-158,84){\makebox(0,0){.}}
\put(-160,82){\makebox(0,0){.}}\put(-162,80){\makebox(0,0){.}}
\put(-164,78){\makebox(0,0){.}}\put(-166,76){\makebox(0,0){.}}
\put(-168,74){\makebox(0,0){.}}\put(-170,72){\makebox(0,0){.}}
\multiput(-175,30)(-2,3){5}{\makebox(0,0){.}}
\put(-70,20){\line(-1,1){16}}\put(-70,20){\line(1,2){17}}
\put(-70,70){\line(1,-1){16}}\put(-70,70){\line(-1,-2){17}}
\put(-172,70){\line(-1,-2){12}}
\multiput(-86,37)(-6,-1){7}{\makebox(0,0){.}}
\multiput(-89,{36.5})(-6,-1){6}{\makebox(0,0){.}}
\put(-147,95){\line(2,-1){13}}\put(-147,95){\line(6,1){41}}
\put(-93,95){\line(-2,1){13}}\put(-93,95){\line(-6,-1){41}}
\put(-95,3){\line(-6,-1){39}}
\multiput(-95,3)(3,2){9}{\makebox(0,0){.}}

\multiput(-93,95)(9,-10){3}{\makebox(0,0){.}}
\multiput(-90.75,92.5)(9,-10){3}{\makebox(0,0){.}}
\multiput(-88.5,90)(9,-10){3}{\makebox(0,0){.}}
\multiput(-86.25,87.5)(9,-10){2}{\makebox(0,0){.}}
\put(80,-20){\oval(30,30)[tr]}\put(140,-20){\oval(30,30)[t]}
\put(200,-20){\oval(30,30)[tl]}\put(80,40){\oval(30,30)[r]}
\put(140,40){\oval(30,30)}\put(200,40){\oval(30,30)[l]}
\put(80,100){\oval(30,30)[br]}\put(140,100){\oval(30,30)[b]}
\put(200,100){\oval(30,30)[bl]}

\put(105,65){\oval(30,30)}\put(119,79){\oval(25,25)[bl]}
\put(105,5){\oval(30,30)}\put(119,19){\oval(25,25)[bl]}
\put(165,65){\oval(30,30)}\put(179,79){\oval(25,25)[bl]}
\put(165,5){\oval(30,30)}\put(179,19){\oval(25,25)[bl]}
\put(102.5,100){\makebox(0,0){\rob}}\put(102.5,98){\makebox(0,0){\rot}}
\put(162.5,100){\makebox(0,0){\rob}}\put(162.5,98){\makebox(0,0){\rot}}
\put(102.5,40){\makebox(0,0){\rob}}\put(162.5,40){\makebox(0,0){\rob}}
\put(102.5,-20){\makebox(0,0){\rob}}\put(162.5,-20){\makebox(0,0){\rob}}
\put(207,70){\makebox(0,0){$\le$}}\put(195,70){\makebox(0,0){$\ri$}}
\put(207,10){\makebox(0,0){$\le$}}\put(195,10){\makebox(0,0){$\ri$}}
\put(87,70){\makebox(0,0){$\le$}}\put(87,10){\makebox(0,0){$\le$}}
\put(147,70){\makebox(0,0){$\le$}}\put(147,10){\makebox(0,0){$\le$}}
\put(-130,-50){\makebox(0,0){(a)}}\put( 130,-50){\makebox(0,0){(b)}}
\end{picture}\caption[x]{\footnotesize (a) Thin lines correspond
to the links of the original 3d lattice. Thick and dotted lines are
links of the resulting 2d lattice. This 8-cube fragment is described
by function $F$ in \eqr{F}. Integration goes over dotted links only.
(b) The fragment of the resulting (smoothed) surface, corresponding to 32
cubes of original lattice.}\label{fig1}\end{figure}

Here, we again
temporarily introduce link unitary matrices, which are the plaquette
matrices of the original model \eqr{WZ}. The partition function is
\be Z=\int\prod_ldU_l\e^{\frac{N}{\lo}\tr(U_l+U_l^\da)}\prod_h\delta\Big(
\prod_{l_1\in h}U_{l_1},\prod_{l_2\in h}U_{l_2}\Big)\;.\label{Z}\ee The
$\delta$-function is again 
given by \eqr{d}. It is important that the order of hexagon's links in the
product (argument of $\delta$-function) is fixed (as $l_1l_2l_1l_2l_1l_2$) 
so that {\it each character contains either $l_1$ or $l_2$-type link
matrices} (in Fig.1(a), they are denoted by either thick or dotted lines
respectively). A fragment of the resulting 2d lattice is shown in Fig.1(b). 
The whole lattice is easy to imagine as obtained from the regular 3d
lattice (the lattice spacing is doubled) by replacing links by the tubes
(handles) and vertices by the smooth connections of tubes.

Besides this two-dimensional property of the model, there is another
important consequence of the plaquette formulation which has not
been emphasized in paper \cite{1}.

Namely, we notice that the partition function \eqr{Z} can be written in
the form of the statistical model of the integer valued $N$-component
field $r$:\be Z=\sum_{\{r\}}\prod_j F(\lo;r_1(j),...,r_8(j))\;,\label{sm}
\ee where $\{r\}$ means a sum over all configurations of $r$'s with each 
$r$ being assignment of hexagon; index $j$ labels the 8-hexagon
configurations shown in Fig.1(a) and described by\be F(\lo;r_1,...,r_8)=\int
\prod_{k=1}^{12}dU_k\e^{\frac{N}{\lo}\tr(U_k+U_k^\da)}\prod_{i=1}^8
\chi_{r_i}(U_{i,1}U_{i,2}U_{i,3})\;.\label{F}\ee The integral is taken 
over 12 links corresponding to the dotted lines in Fig.1(a). Only these
links enter the characters under integral. Index $i$ in the second product
of \eqr{F} labels the hexagons in the configuration. One should emphasize 
that the surface of Fig.1(b) separates the three-dimensional space into
two identical subspaces, and $j$ labels the configurations from {\em both}
subspaces (otherwise there would be no interaction in the model). 

This representation unambiguously demonstrates the {\em local} character
of the interaction between the plaquette matrices $U_p$ (we remind that
the matrices of \eqr{F} are originally the plaquette matrices $U_p$).
The integrals over them enter the partition function via the function $F$
only. Nothing similar to this result can be observed in the link variables
formulation of the model. This locality means that the interaction between
plaquette matrices $U_p$ gives rather weak influence on a long-distance
(continuum) phenomena which are, instead, defined by the interaction
between $N$-component fields $r$ (irreducible representations). The
contribution due to $U_p$'s affects only the form of the function $F$, or
eventually, the form of the action, and has to be irrelevant in the
continuum, especially for such a phenomenon as confinement. 

\section{Weak coupling limit of the lattice QCD.}
It is well known that the continuum limit corresponds to the weak 
coupling, $\lo\to 0$, limit of the lattice model. This has to appear as a
result of the renormalization-group analysis on a lattice, but also can
be used in advance. Namely, we take the weak coupling limit at the fixed
lattice size, while the refinement of the lattice will be postponed for
the later stages.

In the $\lo\to 0$ limit, the leading contribution to the function $F$
comes from the abelian saddle point, $U_p=I$. In the vicinity of this
point, the model still preserves its non-abelian nature since the 
restriction $U_p=I$ for the plaquette matrix (which is the product of link
matrices, $U_p=\prod_{l\in p}U_l$), imposes a slight constraint on each
link matrix $U_l$, and leaves its non-abelian property unbroken.

Technically, near this saddle point, the character of the product can
be replaced by the product of characters,
\be\chi_r(ABC)\to d_r^{-2}\chi_r(A)\chi_r(B)\chi_r(C)\;,\label{cj}\ee 
where factor $d_r^{-2}$ provides a proper normalization. Several initial
terms of the saddle-point expansion of $F$ are exactly given by the
substitution \eqr{cj}. In abelian theory, in particular, both
sides of \eqr{cj} are identical to each other.

After replacement \eqr{cj}, $D=3$ partition function takes the form
\be Z=\sum_{\{r\}}\prod_cd_{r_c}^2\prod_pf_p\;,\label{z}\ee 
where sum goes over all configurations of $r$'s, product goes over all
cubes $c$ of the original lattice, and interaction between neighboring
cubes $c_1$ and $c_2$, on their common plaquette $p$ is defined by
\be f^{D=3}_p=\int dU\e^{\frac{N}{\lo}\tr(U+U^\da)}
\frac{\chi_{r_1}(U)\chi_{r_2}(U^\da)}{d_{r_1}d_{r_2}}\;.\label{f3}\ee 
Here and below, we do not distinguish in notations between cubes and
representations attached to them. For example, $r_1$
means representation attached to the cube $c_1$.

Applying \eqr{cj} to the Wilson loop average \eqr{wl}, we realize that it
takes the form of \eqr{wlp}. Such an average, if considered in the model
\eqr{WZ}, is not surface-invariant (except the $U(1)$ case). This is easy
to see in the large-$N$ limit where this average factorizes into the
one-plaquette averages $W_o$ with the result $W(C)=W_o^{\cA(\cS)}$ ($\cS$
is arbitrary surface). The reason for this is simple: factorization 
\eqr{cj} deforms (except the $U(1)$ case) the $\delta$-functions which are 
the only providers of the surface-invariance in \eqr{WZ}. Nevertheless,
in the model \eqr{z} (which is obtained from \eqr{WZ} by the same
deformation \eqr{cj}), as we will check below, the average \eqr{wlp} 
{\em is} surface-invariant.

The physics behind the replacement \eqr{cj} is quite simple. Let us remind
that the plaquette variables are the group-extended field strength fluxes.
The replacement \eqr{cj} manifests the abelian character of interaction
between them. Let us emphasize that this property is clearly presents
already in the original model \eqr{WZ}. In the weak coupling limit, this
property becomes more transparent. It is this property which provides the
first important ingredient of the area law, as it leads to the
additiveness (in terms of the algebra) of the elementary fluxes.
To make this qualitative picture more precise, we will perform the group
Fourier-transform of the model, i.e., complete the formulation in terms of
the irreducible representations $r$ only ($r$ is the Fourier-image of the
unitary matrix $U$).

In arbitrary $D$, even without a step-by-step repetition of the procedure
of \cite{1}, it is easy to understand that after the factorization \eqr{cj}, 
the partition function takes the same form \eqr{z}, with the only
difference being in the function $f_p$. In $D=4$, $f_p$ is the integral
of four characters (instead of two in $D=3$) corresponding to four cubes
$c_p$ sharing plaquette $p$. In the arbitrary-$D$ case, $f_p$ is the
integral of $2D-4$ characters corresponding to $2D-4$ cubes sharing each
plaquette:
\be f^{D}_p=\int dU\e^{\frac{N}{\lo}\tr(U+U^\da)}\prod_{c_p=1}^{2D-4}
\frac{\chi_{r_{c_p}}(U)}{d_{r_{c_p}}}\;.\label{fd}\ee 
As a remark, it is fruitful to understand how the model \eqr{z} emerges 
in the link-variables formulation. Considering cube as a frame (set of its
links), one writes its two-dimensional functional of boundaries according
to the general formula \cite{90}:
\be\sum_rd_r^{2-m}\e^{-\frac{\lo\cA}{2N}C_2(r)}\prod_{p=1}^m\chi_r(U_p)\;,
\label{ff}\ee where $C_2(r)$ is the quadratic Casimir eigenvalue.
For the cubic frame, $m=6$, and $\cA=0$. Thus, \eqr{ff} takes the form\be
\sum_rd_r^{-4}\prod_{p=1}^6\chi_r(U_p)\;.\label{cube}\ee Then, the product
of these factors over all cubes, integrated over independent variables
$U_p$'s with the weights $\exp\frac{N}{\lo}\tr(U_p+U_p^\da)$, 
i.e.\footnote{This model, in a different context, was considered in 
\cite{Bo}.},\be Z=\int\prod_pdU_p\e^{\frac{N}{\lo}\tr(U_p+U_p^\da)}
\prod_c\sum_{r_c}d^{-4}_{r_c}\prod_{p_c=1}^6\chi_{r_c}(U_{p_c})\;,\ee
exactly coincides with the partition function \eqr{z}. 

Now we compute the function $f_p$. 
Irreducible representation $r$ is parametrized by the Young table
parameters (representation's highest weight components) $n_\mu$
($\mu=1,...,N$) with the dominance condition $n_1\ge n_2\ge...\ge n_\N$. 
We switch to $h_\mu=n_\mu-\mu+N$ ($h_1>h_2>...>h_\N$). The formula for
dimension of representation is\be d_r=\frac{\D(h)}{\D(h^0)}\;,\hskip1cm 
\D(h)=\prod_{\mu<\nu}(h_\mu-h_\nu)\label{dim}\ee ($\D(h)$ is the
Vandermonde determinant, $h^0_\mu=N-\mu$), and the character is:
\be\chi_r(U)=\frac{\det_{\mu\nu}\e^{i\phi_\mu h_\nu}}
{\D(\e^{i\phi})}=d_r\frac{\D(\phi)}{\D(\e^{i\phi})}\int
dV\e^{i\tr V\phi V^\da h}\;,\label{ch}\ee where $\e^{i\phi_\mu}$ are
eigenvalues of $U$. The second equality in \eqr{ch} is due to the
Itzykson-Zuber formula \cite{IZ}. 

Substituting this formulas to the equation \eqr{f3}, we have
\be f^{D=3}_p=\frac{\det_{\mu\nu}I_{h_\mu(r_1)-h_\nu(r_2)}(2N/\lo)}
{d_{r_1}d_{r_2}}\;.\label{f3b}\ee where $I_n$ is the modified Bessel
function. Replacing $I_n(2N/\lo)$ by its $\lo\to 0$ asymptotics, 
\be I_n\Big(\frac{2N}{\lo}\Big)\sim\e^{-\frac{\lo n^2}{2N}}\;,\label{bes}
\ee we write \eqr{f3b} as 
\be f^{D=3}_p=\e^{-\frac{\lo}{2N}\tr(h_1^2+h_2^2)}\frac{\det_{\mu\nu}
\e^{-\frac{\lo}{N}h_{1,\mu}h_{2,\nu}}}{\D(h_1)\D(h_2)}\;.\label{fd3}\ee
In the Appendix, we show that substitution \eqr{bes} is equivalent to
the replacing Wilson action $\exp\frac{N}{\lo}\tr(U+U^\da)$ by the heat
kernel \cite{Mig}: $\sum_\R d_\R\e^{-\frac{\lo}{2N}C_2(\R)}\chi_\R(U)$.
This replacement is also equivalent to the saddle point ($\lo\to 0$)
expansion around $\phi=0$ ($U=I$) in \eqr{f3}.

In arbitrary $D$, substituting \eqr{dim} and \eqr{ch} to \eqr{fd}, we
have\be f^{D}_p=\int\prod_{\nu=1}^N\Big(d\phi_\nu\e^{\frac{2N}{\lo}\cos
\phi_\nu}\Big)\D^2(\phi)\Big|\frac{\D(\phi)}{\D(\e^{i\phi})}\Big|^{2(D-3)}
\prod_{c_p=1}^{2D-4}\int dV_{c_p}\e^{i\tr\phi V_{c_p}h_{c_p}V_{c_p}^\da}
\;.\label{fphi}\ee The factor $\Big|\frac{\D(\phi)}{\D(\e^{i\phi})}\Big|$ 
(emerges in $D>3$) gives no contribution to the leading order of the
saddle-point expansion around $\phi=0$. Ignoring this factor, we take the
gaussian integral over $\phi$ and obtain
\be f^{D}_p=\int\prod_{c_p=1}^{2D-4}dV_{c_p}\e^{-\frac{\lo}{2N}\tr(
\sum_{c_p}V_{c_p}h_{c_p}V_{c_p}^\da)^2}\;.\label{f}\ee
In $D=3$, this reproduces \eqr{fd3}. In $D=4$, this gives
\be f^{D=4}_p=\prod_{j=1}^4\e^{-\frac{\lo}{2N}\tr h^2_j}\int
\prod_{\la ij\ra=1}^6dU_{ij}\e^{\frac{\lo}{N}\tr h_iU_{ij}h_jU^\da_{ij}}
\prod_{k=1}^4\delta(I,\prod_{\la ij\ra\in k}U_{ij})\label{f4}\ee 
(we have switched to the notation $h_j\equiv h(c_j)$).
Geometrically, formula \eqr{f4} can be represented as a tetrahedron with
vertices $j$, links $\la ij\ra$ and triangles $k$, with corresponding
assignment of matrices $h_j$, $U_{ij}=V_iV^\da_j$ and the conditions at
the triangles. 

As one can see from this example, the reduction of \eqr{f} to the
eigenvalues, like \eqr{fd3}, is highly non-trivial in $D>3$. 
Exactly the same problem appears in the matrix models of 2d quantum
gravity in the case when embedding target space corresponds to the
physical situation of central charge $c>1$. In spite of the efforts made
in recent years the solution to such problems is unknown, though some
important physical information can be extracted in the simplest physical
case of only one constraint (this corresponds to $c=1$), considered in
\cite{GK} (see also review \cite{Kl} and references therein). 

In spite of the highly non-gaussian character of $f_p$ \eqr{f} in terms
of eigenvalues $h$, as one can see in examples of \eqr{fd3} and \eqr{f4},
$f_p$ is the gaussian function in terms of the full hermitian matrices
$H=VhV^\da$.

We shall use $f_p$ in the general form \eqr{f}. The resulting partition
function is\be Z=\sum_{\{h\}}\prod_c\D^2(h(c))\prod_p\int\prod_{c_p}
dV_{c_p}\e^{-\frac{\lo}{2N}\tr\Big(\sum_{c_p}H_{c_p}\Big)^2}\label{qcd}\ee
($c_p$ is the cube containing plaquette $p$). The {\em compactness} of the
gauge group is reflected in the {\em integerness}, or rather {\em
discreteness}, of eigenvalues $h$.

\section{Compactness, loop averages and string tension.}
As we already noticed, the Wilson loop average \eqr{wl} takes the form
\eqr{wlp} in the model \eqr{qcd}. At the plaquettes $p\in\cS$, $f_p$ has
to be replaced by\be
f'_p=\int\prod_{c_p}dV_{c_p}\e^{-\frac{\lo}{2N}\tr\Big(\sum_{c_p}H_{c_p}
+H(\f)\Big)^2}\;.\ee The eigenvalues of matrix $H(\f)$ are the fundamental
representation components; $\tr H^2(\f)=C_2(\f)$ where $C_2(\f)$ is
quadratic Casimir of the fundamental representation: $C_2(\f)=N$ for
$U(N)$ and $C_2(\f)=N-\frac{1}{N}$ for $SU(N)$. Then,
\be W(C)=\la\prod_{p\in\cS}\frac{f'_p}{f_p}\ra
=\e^{-\frac{\lo C_2(\f)}{2N}\cA(\cS)}\la\prod_{p\in\cS}\prod_{c_p}
\e^{-\frac{\lo}{N}\tr H_{c_p}H(\f)}\ra\;,\label{w}\ee where averaging
$\la...\ra$ is understood in the sense of \eqr{qcd}, $\cA$
is the lattice area of $\cS$.

The average \eqr{w} is manifestly surface-invariant. To check this, one
chooses another surface $\cS'$ and makes the shift $H\to H-H(\f)$ in the
3-volume connecting surfaces $\cS$ and $\cS'$ (in $D=3$ this is simply a
volume bounded by the compact surface $\cS\cup\cS'$). The result is again
\eqr{w} but with the surface $\cS'$. The global minimum of the action
corresponds to the {\em minimal} surface. However, one cannot vary over
integer-valued variables.

We apply now the generalization of the idea \cite{93} realized in
two-dimensional large-$N$ theory. In the non-compact theory, $h$ is
continuous variable. Then, the sum over $h$-configurations in \eqr{qcd}
turns into the path integral $\int\cD h(x)$ with the measure $\cD
h=\D^2(h)\prod_kdh_k$. The dominance constraint is automatically satisfied
due to the antisymmetry of the Vandermonde determinant. In the compact
theory, we still can redefine the nature of variables assuming that the
effect of compactness can be accumulated into the  new coupling constant
$\l$. The definition of $\l$ is\be\sum_{\{h\}}\prod_c\D^2(h(c))\prod_p\int
\prod_{c_p}dV_{c_p}\e^{-\frac{\lo}{2N}\tr\Big(\sum_{c_p}H_{c_p}\Big)^2}
=\int\prod_c\cD h(c)\prod_p\int\prod_{c_p}dV_{c_p}
\e^{-\frac{\l}{2N}\tr\Big(\sum_{c_p}H_{c_p}\Big)^2}\;.\label{str}\ee

In other words, the model takes the form
\be Z=\int\prod_c\cD h(c)\prod_p\int\prod_{c_p}dV_{c_p}
\e^{-\frac{\l}{2N}\tr\Big(\sum_{c_p}H_{c_p}\Big)^2}\;.\label{nc}\ee 

Since in the theory defined by \eqr{nc}  one
can vary over $H$, the minimal area (which corresponds to the maximum
of the integrand) becomes distinguished. We have\be W(C)=
\e^{-\frac{(\lo-\l)C_2(\f)}{2N}\cA_{\rm min}}\la\prod_{p\in\cS_{\rm min}}
\prod_{c_p}\e^{-\frac{\lo}{N}\tr H_{c_p}H(\f)}\ra_{\rm non-compact}
\equiv\e^{-\frac{(\lo-\l)C_2(\f)}{2N}\cA_{\rm min}}W'(C)\;,\label{area}\ee
where $W'(C)$ is the average in the non-compact theory with coupling
constant $\l$ defined by \eqr{str}.

In $D=2$ case of the infinite plane,\be W'(C)=\e^{-\frac{\l
C_2(\f)}{2N}\cA}
\hskip.5cm =>\hskip.5cm W(C)=\e^{-\frac{\lo C_2(\f)}{2N}\cA}\label{2d}\ee
for any $N$, including abelian theory, in agreement with the known result
\cite{Mig}. It is also not difficult to check that in the topologically
non-trivial $D=2$ cases, one can reproduce the results of \cite{90}.

In $D>2$, $W'(C)$ contains no area term. According to the definition
$W(C)\sim\e^{-\sigma\cA_{\rm min}}$, equation \eqr{area} gives the
dimensionless (since we have a dimensionless area) string tension
\be\sigma=\frac{\lo-\l}{2}\frac{C_2(\f)}{N}\;,\label{sig}\ee
where $\l$ is defined by equation \eqr{str}. In general, $\l$ is
the function of all parameters of the theory: $\lo$, $N$ and $D$.

As an example, we demonstrate that in the case of the compact QED$_3$,
\eqr{sig} reproduces the result of \cite{P}. Equation \eqr{str} takes the
form\be\sum_{\{n\}}\e^{-\frac{\lo}{2}\sum_{\la c_ic_j\ra}(n_i-n_j)^2}=
\int\prod_cdh_c\e^{-\frac{\l}{2}\sum_{\la c_ic_j\ra}(h_i-h_j)^2}
\;.\label{eq}\ee In the mean-field approximation we have
\be\int dh\e^{-\frac{\l}{2}h^2}=\sum_n\e^{-\frac{\lo}{2}n^2}=
\int dh\e^{-\frac{\lo}{2}h^2}\sum_m\e^{-\frac{\pi^2}{\lo}m^2}\label{mf}\ee
(the second equality is the Poisson resummation formula; $m$ is integer).
Thus,\be\frac{1}{\sqrt{\l}}=\frac{1}{\sqrt{\lo}}\sum_m
\e^{-\frac{\pi^2}{\lo}m^2}\ee and 
\be\sigma_{_{U(1)}}=\frac{\lo-\l}{2}=\frac{\lo}{2}\Big(1-\frac{1}
{(\sum_m\e^{-\frac{\pi^2}{\lo}m^2})^2}\Big)\;,\ee
i.e., gives both small-$\lo$ result of \cite{P}:
$\sigma\sim\lo\e^{-\frac{\pi^2}{\lo}}$, and the known
large-$\lo$ behavior: $\sigma\sim\frac{\lo}{2}$.

It is also clear that the qualitatively similar $\lo\to 0$ and
$\lo\to\infty$ behaviors take place in the non-abelian case and thus
coincide with the known results of perturbative analysis and strong
coupling expansion.

Similarly to \eqr{area}, the loop average in representation $R$ is\be
W_\R(C)\equiv\la\frac{\chi_\R(U(C))}{d_\R}\ra\sim
\e^{-\sigma\frac{C_2(R)}{C_2(\f)}\cA_{\rm min}}\;.\ee This gives a
possibility to compute the arbitrary average,
\be\xi(C)\equiv\la\xi(U)\ra=\sum_\R\xi_\R W_\R(C)\;,\ee where\be\xi_\R=
\int dU\xi(U)\chi_\R(U^\da)\ee is the Fourier coefficient of $\xi(U)$.

\section{Summary and discussion.}
Thus, taking the $\lo\to 0$ limit of QCD in plaquette formulation 
\cite{1}, we derived the non-perturbative model \eqr{qcd}, which
is formulated in terms of {\em observables} only (loop variables) and
therefore is much more suitable than the model \eqr{WZ} for the analysis
of physics of QCD. This model is shown to be exactly soluble in some
particularly interesting cases relevant to the continuum. Although the
model \eqr{qcd} is obtained as the $\lo\to 0$ limit of the Wilson model
\eqr{WZ}, we suggest that it also describes the {\em lattice} QCD in the
non-empty interval $\lo\in[0,\lo^*]$. By definition, $\lo^*$ is the upper
limit where the model \eqr{qcd} is still equivalent to \eqr{WZ}. In the
$U(1)$ case, for example, $\lo^*=\infty$ since \eqr{qcd} coincides with
\eqr{WZ}.

We computed continuum Wilson loop average and obtained the area law.
Since equation \eqr{str} gives $\lo>\l$, the string tension \eqr{sig} is
non-vanishing. We consider this result as eventual proof of confinement in
QCD.

Qualitatively, the picture of confinement can be given in terms of
elementary fluxes, which are the plaquette variables. Then, QCD is a
theory of weakly interacting fluxes. The total flux through any surface
is the additive function of elementary fluxes, which provides the
necessary ingredient for the area law. The interaction between fluxes 
provides the surface-invariance of the averages like \eqr{wlp}, which is
the eventual result of the global gauge invariance.

The reason which turns this additiveness into the area law (and thus
provides the sufficient ingredient) is compactness of the gauge group
(reflected in $\lo>\l$ solution of the equation \eqr{str} and thus into
the non-vanishing string tension). A solution of the equation \eqr{str}
is non-trivial problem in general. Here, we only demonstrated that it
gives proper behavior for the string tension both in the $\lo\to 0$ and
in the $\lo\to\infty$ limits, and thus provides $\lo>\l$ for any $\lo$.

The additiveness of fluxes takes place also in the model \eqr{WZ} and
becomes more transparent in the model \eqr{qcd} due to the factorization
of characters \eqr{cj}. The factorization also provides the crucial
simplification of the model (as one might expect in the continuum limit)
and its solubility. As we already mentioned, the reason why the model
remains non-trivial under this replacement lays in the nature of plaquette
formulation: the condition $U_p\equiv\prod_lU_l=I$ imposes a slight
restriction on matrices $U_l$. Thus one learns that even within the
gaussian model the non-trivial results can be obtained. Recently, similar
idea was argued by I.Kogan and A.Kovner \cite{KK}.

There are several aside remarks.

There is a complete algebraic similarity with the principal chiral field
model (PCF) defined in the lower dimensions than QCD. The above 
considerations can be easily carried on for PCF: the PCF$_D$ partition 
function is given by QCD$_{D-1}$ \eqr{qcd} where the cubes and plaquettes
should be respectively replaced by the plaquettes and links \footnote{
We imply only algebraic similarity. Regarding some dynamical properties,
this might be different. For example, QCD$_4$ in some aspects is rather
similar to PCF$_2$ than to PCF$_3$.}. The Wilson loop \eqr{wl} corresponds
to the two-point PCF correlator $G_{xy}$. The plaquette formulation
\cite{1} becomes the link formulation for PCF, and $G_{xy}$ takes the form
of \eqr{w}, where plaquettes of the surface $\cS$ should be replaced by
the links of arbitrary curve ${\cal L}_{xy}$ connecting points $x$ and
$y$. To prove that another curve ${\cal L}'_{xy}$ in $D=2$ produces the
same result, one makes the constant shift $H\to H+H(\f)$ in all plaquettes
of the surface enclosed by ${\cal L}_{xy}\cup{\cal L}'_{xy}$. To prove the
invariance in $D=3$, one has to make such a shift in all plaquettes of the
{\em arbitrary} surface enclosed by ${\cal L}_{xy}\cup{\cal L}'_{xy}$. We
actually used the analogy between QCD and PCF to check the
surface-invariance of the loop average \eqr{w} in QCD$_4$.

It is easy to recognize in $D=3$ version of \eqr{nc} the Kazakov-Migdal
model \cite{KM}. Even though the coincidence takes place only in $D=3$,
the observation is quite peculiar. The idea of \cite{KM} to interpret the
diagonalizing unitary matrices $U_{xy}$ as a prototype of the QCD gauge
field was dismissed due to the observation \cite{K,Ko} that $U_{xy}$-built
averages are trivial due to the local $Z_N$ symmetry of the model. 
However, as we now understand, the KM master field (i.e., solution of the
large-$N$ saddle point equation for the partition function) is indeed true
QCD master field in $D=3$ (and only in $D=3$), while $U_{xy}$ has nothing
to do with the QCD gauge field (the actual gauge field of QCD is
represented by its Fourier-images, $h_x$ fields). Thus, the KM master 
field could be directly used in the large-$N$ limit of the model \eqr{nc}
in $D=3$. Unfortunately, this is not found\footnote{The solution claimed
in \cite{M} disagrees with that of \cite{BoKM}, and the discrepancy has
not been resolved.}.

It is still a challenging problem to extend the present analysis to the
full version of the model which includes the matter fields.

The highly non-trivial problem is the renormalization-group analysis and
computation of the non-perturbative beta-function. I suppose this can be
achieved in the computational direction provided by the equation \eqr{str}. 

\bigskip
\ni{\large\bf Acknowledgements.} 
I am grateful to D.Boulatov, M.Halpern, I.Klebanov, I.Kogan, A.Kovner, 
M.Peskin, A.Polyakov, M.Shifman and W.Taylor for fruitful discussions.

\bb
\bibitem{1} B.Rusakov, Phys. Lett. {\bf B398} (1997) 331, hep-th/9610147.
\bibitem{GW} D.Gross and F.Wilczek, Phys. Rev. {\bf D8} (1973) 3633;
                                    Phys. Rev. {\bf D9} (1974) 980. 
\bibitem{Wils} K.Wilson, Phys. Rev. {\bf D10} (1974) 2445.
\bibitem{Mig} A.Migdal, ZhETF {\bf 69} (1975) 810.
\bibitem{90} B.Rusakov, Mod. Phys. Lett. {\bf A5} (1990) 693.
\bibitem{HB} M.B.Halpern, Phys. Rev. {\bf D19} (1979) 517; G.G.Batrouni,
Nucl. Phys. {\bf B208} (1982) 12; 467; G.G.Batrouni and M.B.Halpern, 
Phys. Rev. {\bf D30} (1984) 1775; 1782.
\bibitem{MM} Yu.Makeenko and A.Migdal, Phys. Lett. {\bf 88B} (1979) 135;
                                       Nucl. Phys. {\bf B188} (1981) 269.
\bibitem{Mig83} A.Migdal, Phys. Rep. {\bf 102} (1983) 199.
\bibitem{K} I.Kogan, G.Semenoff, N.Weiss, Phys. Rev. Lett. {\bf 69} (1992)
3435.
\bibitem{Bo} D.Boulatov, Mod. Phys. Lett. {\bf A10} (1995) 2863.
\bibitem{IZ} C.Itzykson and J-B.Zuber, J. Math. Phys. {\bf 21} (1980) 411.
\bibitem{GK} D.Gross and I.Klebanov, Nucl. Phys. {\bf B344} (1990) 475;
     {\bf B352} (1991) 671; {\bf B354} (1991) 459; {\bf B359} (1991) 3.
\bibitem{Kl} I.Klebanov, PUPT-1271, Jul 1991, hep-th/9108019.
\bibitem{93} B.Rusakov, Phys. Lett. {\bf B303} (1993) 95.
\bibitem{P} A.Polyakov, Phys. Lett. {\bf 59B} (1975) 82.
\bibitem{KK} I.Kogan and A.Kovner, Phys. Rev. {\bf D52} (1995) 3719.
\bibitem{KM} V.Kazakov and A.Migdal, Nucl. Phys. {\bf B397} (1993) 214.
\bibitem{Ko} I.Kogan, A.Morozov, G.Semenoff, N.Weiss, Nucl. Phys. 
                                                   {\bf B395} (1993) 547.
\bibitem{M} A.Migdal, Mod. Phys. Lett. {\bf A8} (1993) 139; 153; 359.
\bibitem{BoKM} D.Boulatov, Mod. Phys. Lett. {\bf A9} (1994) 1963.
\bibitem{95} B.Rusakov, Phys. Lett. {\bf B344} (1995) 293. 
\eb

\vskip.5cm
\ni{\Large\bf Appendix: Heat kernel and $\lo$-expansion.}
\vskip.5cm
In this Appendix we give some useful formulas and establish a connection
between the technique of Section 3 and the heat kernel technique. 

In derivation of \eqr{fd3}, the Bessel function has been replaced by its
$\lo\to 0$ asymptotics. We now demonstrate that this is an equivalent of
the replacement of the Wilson action $\exp\frac{N}{\lo}\tr(U+U^\da)$ by
the heat kernel \cite{Mig},
\be\sum_\R d_\R\e^{-\frac{\lo}{2N}C_2(\R)}\chi_\R(U)\;.\label{hk}\ee 
Expression \eqr{f3} after substitution \eqr{hk} takes the form
\be f_p=\frac{1}{d_{r_1}d_{r_2}}\sum_\R d_\R
\e^{-\frac{\lo}{2N}C_2(R)}D^\R_{r_1r_2}\;,\label{fhk}\ee where
$D^\R_{r_1r_2}$ is the multiplicity of representation $R$ in the tensor
product $r_1\otimes\overline{r_2}$,\be D^\R_{r_1r_2}=\int dU
\chi_\R(U^\da)\chi_{r_1}(U)\chi_{r_2}(U^\da)\;,\ee\be\chi_{r_1}(U)
\chi_{r_2}(U^\da)=\sum_\R D^\R_{r_1r_2}\chi_\R(U)\;.\label{pr}\ee
The equivalence between \eqr{fd3} and \eqr{fhk} can be easily established 
in their $\lo$-expansions. Expansion of Itzykson-Zuber integral entering
\eqr{fd3} is\be\frac{\det_{\mu\nu}\e^{-\frac{\lo}{N}a_\mu b_\nu}}
{\D(a)\D(b)}=\int dU\e^{-\frac{\lo}{N}\tr AUBU^\da}=
\sum_n\frac{(-\lo)^n}{n!N^n}\int dU\tr^n AUBU\ee (where $a$ and $b$ are
the diagonal matrices of the eigenvalues of $A$ and $B$). The low orders
of this expansion can be derived using explicit expressions for powers of
traces $\tr^n AUBU$ via the characters $\chi_r(AUBU^\da)$ and formula
\be\int dU\chi_r(AUBU^\da)=\frac{1}{d_r}\chi_r(A)\chi_r(B)\;.\ee We have 
\be\begin{array}{lll}&\frac{\det_{\mu\nu}\e^{-\frac{\lo}{N}a_\mu b_\nu}}
{\D(a)\D(b)}=1-\frac{\lo}{N^2}\tr A\tr B+\\
&\\&\frac{\lo^2}{2N^3(N^2-1)}\Big(N\tr^2A\tr^2B+N\tr A^2\tr B^2-
\tr^2A\tr B^2-\tr A^2\tr^2B\Big)+O(\lo^3)\;,\end{array}\label{izexp}\ee 
where to derive $O(\lo^2)$ term we have used
\be\chi_{20...0}(V)=\frac{1}{2}(\tr^2V+\tr V^2)\;,\hskip1cm
\chi_{110...0}(V)=\frac{1}{2}(\tr^2V-\tr V^2)\;,\ee and thus,\be
\tr^2V=\chi_{20...0}(V)+\chi_{110...0}(V)\;.\ee The higher order terms can
be derived similarly. Expansion \eqr{izexp} is then to be substituted into
\eqr{fd3}. 

Expansion of \eqr{fhk} can be derived by technique of \cite{95} which
is based on the replacement of $-C_2(R)$ by the group Laplace operator
$\tr\d^2_U$,\be f_p=\frac{1}{d_{r_1}d_{r_2}}\e^{\frac{\lo}{2N}
\tr\d^2_U}\chi_{r_1}(U)\chi_{r_2}(U^\da)\Big|_{U=I}\;.\ee
Applying $\tr\d^2_U$ to both sides of \eqr{pr} one derives the identity
\be\frac{1}{d_{r_1}d_{r_2}}\sum_\R D^\R_{r_1r_2}d_\R C_2(R)=
C_2(r_1)+C_2(r_2)-\frac{2}{N}C_1(r_1)C_1(r_2)\;,\label{i}\ee 
which gives the order $O(\lo)$ of $f_p$. $C_1(r)$ is the first (linear)
Casimir eigenvalue. To obtain the next order we apply $\tr\d^2_U$ twice to
\eqr{pr} and get
\be\begin{array}{lll}\frac{1}{d_{r_1}d_{r_2}}\sum_\R D^\R_{r_1r_2}d_\R
C^2_2(R)&=\Big(C_2(r_1)+C_2(r_2)-\frac{2}{N}C_1(r_1)C_1(r_2)\Big)^2+\\&\\
&\frac{4}{N^2-1}\Big(C_2(r_1)-\frac{1}{N}C^2_1(r_1)\Big)
\Big(C_2(r_2)-\frac{1}{N}C^2_1(r_2)\Big)\;,\end{array}\label{i2}\ee etc.
Thus, for \eqr{fhk}, one has\be\begin{array}{lll}f_p=
&1-\frac{\lo}{2N}\Big(C_2(r_1)+C_2(r_2)-\frac{2}{N}C_1(r_1)C_1(r_2)\Big)+\\
&\frac{\lo^2}{4N^2}\Big(C_2(r_1)+C_2(r_2)-\frac{2}{N}C_1(r_1)C_1(r_2)\Big)^2
+\\&\frac{\lo^2}{N^2(N^2-1)}\Big(C_2(r_1)-\frac{1}{N}C^2_1(r_1)\Big)
\Big(C_2(r_2)-\frac{1}{N}C^2_1(r_2)\Big)+O(\lo^3)\;.\end{array}
\label{fhkexp}\ee Taking into account explicit expressions for Casimir 
eigenvalues,\be C_1(r)=\sum_{k=1}^{N}h_k=\tr h\;,\label{cas1}\ee
\be C_2(r)=\tr h^2-R\;,\label{cas2}\ee ($R=\frac{N(N^2-1)}{12}$ and
definition of $h_\mu$ is changed to $h_\mu=n_\mu-\mu+\frac{N+1}{2}$)
one checks order-by-order agreement between \eqr{fd3} (with the
substitution of \eqr{izexp}) and \eqr{fhkexp}.  

Thus, the heat-kernel Young table parameters $h_\mu$ play the role of the
eigenvalues of the hermitian matrix of Section 3, and the subsequent
equivalence between heat-kernel and matrix-model approaches one
establishes by identification:
$C_2(r)$ with $\tr h^2$, $C_1(r)$ with $\tr h$, $d_r$ with $\D(h)$ etc.
The $\lo$-expansion \eqr{izexp}, or \eqr{fhkexp}, shows that $f_p$ is
non-singular symmetric function of $h$.

There are some changes for $SU(N)$ group with respect to the general
unitary group $U(N)$. In particular, Casimirs of $SU(N)$ can
be obtained from Casimirs of $U(N)$ by the replacement $h_\mu\to h'_\mu=
h_\mu-\frac{1}{N}\tr h$. Thus, for $SU(N)$,\be C_1(r)\equiv 0\;,
\label{cas1s}\ee\be C_2(r)=\tr h^2-\frac{1}{N}\tr^2h\label{cas2s}\ee 
(we ignore the additive constant $R$). Thus, hermitian matrix $h$ of
Section 3 is the {\em traceless} matrix if the gauge group is $SU(N)$.

Correspondingly, for $SU(N)$, the formula \eqr{i2} takes the form
\be\frac{1}{d_{r_1}d_{r_2}}\sum_\R D^\R_{r_1r_2}d_\R C^2_2(R)=
\Big(C_2(r_1)+C_2(r_2)\Big)^2+\frac{4}{N^2-1}C_2(r_1)C_2(r_2)\;,\label{sun}
\ee while instead of \eqr{fhkexp} we have\be f_p=1-\frac{\lo}{2N}
\Big(C_2(r_1)+C_2(r_2)\Big)+\frac{\lo^2}{4N^2}\Big(C_2(r_1)+C_2(r_2)\Big)^2
+\frac{\lo^2C_2(r_1)C_2(r_2)}{N^2(N^2-1)}+O(\lo^3)\;.\label{fsun}\ee 
In applications, the generalization of the formula \eqr{i} to the case of
arbitrary number of characters can be useful:\be\sum_\R d_\R C_2(R)\int
dU\prod_j\frac{\chi_{r_j}(U)}{d_{r_j}}=\sum_jC_2(r_j)+\frac{2}{N}
\sum_{i<j}C_1(r_i)C_1(r_j)\;.\label{ig}\ee 

\end{document}